\documentclass[fleqn,usenatbib]{mnras}

\usepackage{newtxtext,newtxmath}

\usepackage[T1]{fontenc}

\DeclareRobustCommand{\VAN}[3]{#2}
\let\VANthebibliography\thebibliography
\def\thebibliography{\DeclareRobustCommand{\VAN}[3]{##3}\VANthebibliography}

\usepackage{graphicx}	
\usepackage{amsmath}	

\usepackage{scalerel,tikz}
\usetikzlibrary{svg.path}
\definecolor{orcidlogocol}{HTML}{A6CE39}
\tikzset{orcidlogo/.pic={
 \fill[orcidlogocol] svg{M256,128c0,70.7-57.3,128-128,128C57.3,256,0,198.7,0,128C0,57.3,57.3,0,128,0C198.7,0,256,57.3,256,128z};
 \fill[white] svg{M86.3,186.2H70.9V79.1h15.4v48.4V186.2z}
 svg{M108.9,79.1h41.6c39.6,0,57,28.3,57,53.6c0,27.5-21.5,53.6-56.8,53.6h-41.8V79.1z M124.3,172.4h24.5c34.9,0,42.9-26.5,42.9-39.7c0-21.5-13.7-39.7-43.7-39.7h-23.7V172.4z}
 svg{M88.7,56.8c0,5.5-4.5,10.1-10.1,10.1c-5.6,0-10.1-4.6-10.1-10.1c0-5.6,4.5-10.1,10.1-10.1C84.2,46.7,88.7,51.3,88.7,56.8z};
}}
\newcommand\orcidicon[1]{\href{https://orcid.org/#1}{\mbox{\scalerel*{
\begin{tikzpicture}[yscale=-1,transform shape]
\pic{orcidlogo};
\end{tikzpicture}
}{|}}}}

\title[Wide-binary eccentricity in star clusters]{Wide-binary eccentricity distribution in young star clusters: dependence on the binary separation and mass}

\author[Mathew et al.]{
Sajay Sunny Mathew$^{\orcidicon{0000-0002-8381-8195}\,1}$\thanks{E-mail: \href{mailto:sajay.mathew@anu.edu.au}{sajay.mathew@anu.edu.au}},
Siyao Xu$^{\orcidicon{https://orcid.org/0000-0002-0458-7828}\,2}$,
Christoph Federrath$^{\orcidicon{0000-0002-0706-2306}\,1}$,
Yue Hu$^{\orcidicon{orcid:0000-0002-8455-0805}\,3}$,
Amit Seta$^{\orcidicon{0000-0001-9708-0286}\,1}$
\\
$^{1}$Research School of Astronomy and Astrophysics, Australian National University, Canberra, ACT~2611, Australia\\
$^{2}$Department of Physics, University of Florida, 2001 Museum Rd., Gainesville, FL 32611, USA\\
$^{3}$Department of Physics, University of Wisconsin-Madison, Madison, WI 53706, USA\\
}

\date{Accepted XXX. Received YYY; in original form ZZZ}

\pubyear{2024}

\begin{document}
\label{firstpage}
\pagerange{\pageref{firstpage}--\pageref{lastpage}}
\maketitle

\begin{abstract}
We study the wide-binary eccentricity ($e$) distribution in young star clusters and the role of turbulence in setting the form of the $e$ distribution using magnetohydrodynamical (MHD) simulations of star cluster formation. The simulations incorporate gravity, turbulence, magnetic fields, protostellar heating, and jets/outflows. We find that (1) simulations that employ purely compressive turbulence driving produce binaries with a superthermal $e$ distribution ($\alpha>1$ in $p(e) \propto e^\alpha$), while simulations with purely solenoidal driving or natural mixture of driving modes produce subthermal/thermal distributions ($\alpha \leq$ 1), (2) the $e$ distribution over the full range of binary separations in our simulations is set at the early stages of the star cluster formation process, (3) while binaries (separation of $r_{\mathrm{pair}} \leq 1000\, \mathrm{AU}$) have subthermal to thermal $e$ distributions ($\alpha \sim 0.8$), wide binaries ($r_{\mathrm{pair}} > 1000\, \mathrm{AU}$) have a superthermal distribution ($\alpha \sim 1.8$), and (4) low-mass binary systems (system masses of $M_{\mathrm{sys}} \leq 0.8\, \mathrm{M_\odot}$) have a highly superthermal distribution ($\alpha \sim 2.4$), whereas high-mass systems ($M_{\mathrm{sys}} > 0.8\, \mathrm{M_\odot}$) exhibit a subthermal/thermal distribution ($\alpha \sim 0.8$). The binary eccentricity distribution is often modelled as a thermal distribution. However, our results suggest that the $e$ distribution depends on the range of separation of the sampled binaries, 
which agrees with the findings from recent {\it Gaia} observations. We conclude that the dependence of the $e$ distribution on the binary separation and mass is linked to the binary formation mechanism governed by the turbulent properties of the parent cloud.  
\end{abstract}




\section{Introduction}
The prevalent existence of binary stars places them at the forefront of many fundamental astrophysical problems 
(e.g., \citealt{Bah85,Mon14,Pena16,Liu19,2022arXiv220700680A,Taur23,2023ASPC..534..275O,Chen24}). As a key parameter of binary orbital dynamics, the eccentricity $e$ is often modelled to follow a thermal\protect \footnotemark[1] eccentricity distribution 
$p(e) = dN/de = 2e$ \citep{Jean19}. Recent {\it Gaia} observations with high-quality parallaxes and proper motions \citep{Gaia23} have significantly expanded the sample of binaries \citep{El21}, including wide binaries with semimajor axes $a\gtrsim 10^3\, \mathrm{AU}$ \citep{2018MNRAS.480.4884E,Hwang22}, which enables detailed statistical studies on their eccentricity distribution. The {\it Gaia} observations indicate a non-universal eccentricity distribution among binary stars in the solar neighbourhood, with a transition from a uniform distribution at separations $\sim 10^2\, \mathrm{AU}$ to a superthermal\protect \footnotemark[1] distribution at separations greater than $10^3\, \mathrm{AU}$ \citep{Hwang22}. This shift challenges our traditional understanding of binary eccentricity distributions which are usually modelled as a thermal distribution. The superthermal eccentricity distribution of wide binaries cannot be accounted for by their evolution in the Galaxy under the effects of Galactic tides and encounters with passing stars \citep{Mod23,Hamil23}. \citet{Hamil23} suggest that the observed superthermal distribution is more likely a result of an even more superthermal initial distribution at the time of their formation. 

The formation mechanisms and statistical properties of binaries, such as eccentricity, in the regime of $a \lesssim 100\, \mathrm{AU}$, have been extensively studied with hydrodynamic and magnetohydrodynamic simulations in the literature \citep[e.g.,][]{1999ApJ...513..252O,2000MNRAS.314...33B,2010ApJ...725.1485O,ryu2017,2019ApJ...887..232L,2023MNRAS.521..866R,kuruwita&haugboulle2023}. For instance, \citet{2014MNRAS.442..285B} find that the $e$ distribution is almost uniform in their star cluster formation simulations comprising of a large population of M-dwarfs and later type stars \citep[see also][]{guszejnov23}. The median binary separation calculated in the simulations of \citet{2014MNRAS.442..285B} is $\sim20\, \mathrm{AU}$, which agrees with the peak of the separation distribution obtained for M-dwarf systems in observational surveys \citep{2012ApJ...754...44J,2019AJ....157..216W}. At similar binary separation ranges, observations also find a uniform eccentricity distribution \citep{tokovinin2016,Hwang22}. With the new finding that the $e$ distribution has a strong dependence on the binary separation \citep{2020MNRAS.496..987T,Hwang22}, which in turn, is dependent on the binary mass \citep{2010ApJS..190....1R,2012ApJ...754...44J,2019AJ....157..216W} and the evolutionary state \citep{2016ApJ...818...73T,2022ApJ...925...39T,kuruwita&haugboulle2023} (see Fig.~\ref{fig:a_all}), it remains to be extensively studied what mechanisms play dominant roles in the origin of binary eccentricity on different mass and separation scales. In particular, the formation mechanisms and statistical  properties of wide binaries with $a \gg 100\, \mathrm{AU}$ need further numerical investigation.  

\footnotetext[1]{A thermal $e$ distribution, $p(e) = 2e$, is proposed to arise when a group of binaries undergo sufficient dynamical interactions with exchange of energy such that the group approximately approaches a state of energy equipartition and follows Boltzmann distribution \citep{Jean19,1937AZh....14..207A,2008LNP...760..181K,Gell19}. $p(e)$ is referred to as a superthermal distribution when $p(e) \propto e^\alpha$ with $\alpha>1$ and as subthermal where $p(e) \propto e^\alpha$ with $\alpha<1$. The distribution is uniform when $\alpha\sim0$ \citep{Hwang22,Mod23}.}

Recent {\it Gaia} observations have new evidence of the important role of turbulence in star and binary formation. Studies by \cite{Ha21,Ha22} have observationally demonstrated that young stars born in turbulent interstellar medium inherit turbulent velocities from their parent cloud. They found that the relative velocities of pairs of young stars are well coupled with the turbulent velocities of the surrounding gas, exhibiting a power-law dependence on their separations across scales within $1-100$~pc.
\citet{Xu23} propose that wide binaries naturally form in turbulent molecular clouds, with the scaling relation between the initial relative velocities and initial separations of binary stars regulated by turbulence. They found this initial condition governed by turbulence naturally results in a superthermal distribution of eccentricities at birth. Using star cluster formation simulations, in this paper, we study the eccentricity distribution of binary star systems and their dependence on the turbulent properties of the parent cloud.

The ubiquitous magnetohydrodynamics (MHD) turbulence in the interstellar medium \citep[e.g.,][]{Arm95,Chep10,Planck16,Hu19,Ha22}
plays an essential role in star and binary formation (e.g., \citealt{Fed12,Krum12,Far24}). In this study, we use MHD turbulence simulations of star cluster formation \citep{2021MNRAS.507.2448M,2023MNRAS.518.5190M} to analyse the eccentricity distribution of binaries within $\sim 1\, \mathrm{Myr}$ of their formation. The simulations incorporate gravity, turbulence, magnetic fields, protostellar radiative heating, and jets/outflows. Including all the above physical mechanisms is essential since each of these mechanisms has a significant effect on the star cluster formation process. Turbulence dictates the gas density distribution of the cloud and hence regulates fragmentation \citep{1994ApJ...423..681V,1997MNRAS.288..145P,2002ApJ...576..870P,2007ApJ...665..416K,2007ApJ...658..423K,2008ApJ...684..395H,2008ApJ...688L..79F,2009ApJ...702.1428H,2012MNRAS.423.2037H,2013MNRAS.430.1653H,2013MNRAS.436.1245F,2015MNRAS.448.3297F,2018MNRAS.480.3916M,2019ApJ...878..157X,2021ApJ...916...83A,2022MNRAS.514..957S}. The bipolar outflows impact the protostellar mass through the removal of accreting material from protostars \citep{2006ApJ...640L.187L,2014ApJ...790..128F,2014prpl.conf..451F,2021MNRAS.507.2448M,2021MNRAS.502.3646G,2024A&A...683A..13L}, and rearrange the gas around these stars, which in turn influences the formation of new stars in their vicinity \citep{2010ApJ...709...27W,2014ApJ...790..128F}. Magnetic fields and protostellar heating provide additional pressure support and suppress fragmentation into multiple systems \citep[e.g.,][]{2011ApJ...740...74K,2017JPhCS.837a2007F,2019FrASS...6....7K,2019ApJ...878..157X,2020MNRAS.496.5201M,2020ApJ...904..194H}. Further, star-forming clouds can have turbulent magnetic fields \citep[e.g.,][]{2017ApJ...847...92H,2018MNRAS.474.5122K,Hu19,2023MNRAS.524.4431H,2024MNRAS.530.1066L} and the fraction of energy in the turbulent component can have important consequences for binary formation \citep{2019MNRAS.485.5532G}.

Our primary goal is to numerically probe the possible origin of the observed non-universal $p(e)$ of binaries with $a \gtrsim 100$ AU and its dependence on binary separation, system mass, and mass ratio. Previous numerical works  
focused on the role played by dynamical interaction and gas friction/gas accretion in the formation of closer binaries ($a < 100\, \mathrm{AU}$) \citep[e.g.,][]{1999ApJ...513..252O,2000MNRAS.314...33B,ryu2017,2019ApJ...887..232L,2023MNRAS.521..866R}. Since the importance of these two formation mechanisms in the regime of close binaries has been already
extensively studied in the literature \citep[e.g.,][]{2009MNRAS.392..590B,2012MNRAS.419.3115B,2019MNRAS.484.2341B}, we focus on the third new mechanism of scale-dependent turbulent velocities of gas inherited by wide binaries ($\gg 100\, \mathrm{AU}$), which has not been studied with numerical simulations before.  We will test the analytical predictions by \citet{Xu23} regarding the formation of wide binary stars in turbulent environments and the origin of their superthermal $p(e)$. Specifically, we will also for the first time investigate the role of turbulence with different driving conditions in shaping $p(e)$. The paper is organized as follows. In Section~\ref{sec: methods}, we detail the simulation data utilised in our study and describe the methodology for calculating the orbital eccentricity of binary systems. In Section~\ref{sec: results}, we perform statistical analysis on $p(e)$ and its dependence on turbulence driving mechanism, binary separation, system mass, and mass ratio. In Section~\ref{sec: discussion}, we delve into the implications of our findings. Section~\ref{sec: limitation} presents the limitations of this study, and in Section~\ref{sec: conclusions}, we summarise the key results.

\section{Methodology}
\label{sec: methods}
To study the properties of young binary stars, we use the star cluster formation simulations from \citet{2021MNRAS.507.2448M} and \citet{2023MNRAS.518.5190M}. These simulations are carried out in a 3D computational domain of $2 \times 2 \times 2\, \mathrm{pc^3}$ dimensions with triply periodic boundaries. An adaptive mesh refinement (AMR) grid structure is used and the maximum effective resolution is $4096^3$ cells with a minimum cell size of approximately $100$ AU. The star cluster formation is modelled by solving the MHD equations with gravity using a modified version of the \textsc{flash} (version~4) code \citep{2000ApJS..131..273F,2008ASPC..385..145D}.

The simulations incorporate a vast array of physical mechanisms including gravity, magnetic fields, turbulence, and protostellar feedback in the form of radiative heating and mechanical outflows. Sink particles are used to model protostellar objects. Whenever a spherical region of radius $r_{\mathrm{sink}}=250\, \mathrm{AU}$ centered on a cell satisfies the criterion for gravitational collapse \citep{2010ApJ...713..269F}, the gas within the region is replaced by a sink particle (see discussion on limitations in \S\ref{sec: limitation}). The position, linear momentum, and angular momentum of the sink particle are determined by the respective quantities of the enclosed gas within the spherical volume of radius $r_{\mathrm{sink}}$ (which is set to be the accretion radius of the sink particle). Sub-grid models are used for modelling the protostellar heating \citep{2020MNRAS.496.5201M} and outflow feedback \citep{2014ApJ...790..128F,2021MNRAS.507.2448M}. We encourage the reader to refer to \citet{2021MNRAS.507.2448M} and \citet{2023MNRAS.518.5190M} for more detailed description of the simulation model used here. 

\subsection{Turbulence driving}
\label{sec:turb}
We drive turbulence within the gas in the computational domain by introducing a forcing term in the momentum equation of MHD. The forcing (acceleration) field is modelled using a stochastic Ornstein-Uhlenbeck (OU) process \citep{1988CF.....16..257E}, which enables us to drive turbulence continuously, with the field varying smoothly in space and time. The turbulence driving module injects kinetic energy on the largest scales, which naturally cascades down to smaller scales, producing a velocity power spectrum $\sim k^{-2}$ or equivalently a velocity dispersion -- size relation of $\sigma_v \propto \ell^{1/2}$ \citep{1981MNRAS.194..809L,2002A&A...390..307O,2004ApJ...615L..45H,2011ApJ...740..120R,2013MNRAS.436.1245F,2021NatAs...5..365F}. Our forcing approach allows us to control the fraction of compressive-to-solenoidal modes by performing a Helmholtz decomposition. When the total power in the forcing is contributed by compressive modes, we obtain a purely compressive driving (curl-free), and when the total power is associated with solenoidal modes, we get a purely solenoidal turbulence driving (divergence-free). When no decomposition is carried out, a natural mixture of driving modes emerges \citep[see][for more details on the turbulence driving method adopted here]{2008ApJ...688L..79F,2010A&A...512A..81F}. A public version of the turbulence driving module is available \citep{2022ascl.soft04001F}.

\subsection{Initial conditions}
\label{sec: initial conditions}
The gas density in the computational domain is initially uniform with $\rho_0 = 6.56 \times 10^{-21}\, \mathrm{g\, cm^{-3}}$. Since the size of the box $L = 2\, \mathrm{pc}$, the total cloud mass $M_{\mathrm{cl}}= \rho_0 \times L^3 \approx 775\, \mathrm{M_{\odot}}$, yielding a mean free-fall time of $t_{\mathrm{ff}} \approx 0.82\, \mathrm{Myr}$. Once the turbulence driving module \citep{2010A&A...512A..81F,2022ascl.soft04001F} is turned on, over-densities in the form of clumps and filaments are created due to the stirring of the gas and the presence of turbulent shocks. The magnetic field, which is uniform initially with $B= 10^{-5}\, \mathrm{G}$ along the z-axis of the computational domain, is modified as a result of the twisting, mixing and compression \citep{2002PhRvL..88x5001C}, and elongation of magnetic field lines by the turbulent motions \citep{2021PhRvF...6j3701S}, generating a field structure similar to that observed in real molecular clouds \citep{2016JPlPh..82f5301F}.

The turbulence driving is first performed without self-gravity for two turbulent crossing times, $2 t_\mathrm{turb}=L/(\mathcal{M}c_\mathrm{s}) = 2\,\mathrm{Myr}$ \citep{2010A&A...512A..81F}, where $\mathcal{M}$ is the steady state sonic Mach number and $c_\mathrm{s}=0.2\,\mathrm{km/s}$ is the isothermal sound speed for solar metallicity, molecular gas at $10\,\mathrm{K}$.  The velocity dispersion on the turbulence driving scale is set as $\sigma_v=c_\mathrm{s}\,\mathcal{M} = 1.0\, \mathrm{km\, s^{-1}}$ such that $\mathcal{M}=5$ and Alfv$\acute{\text{e}}$n Mach number $\mathcal{M_{\mathrm{A}}}=2.9$. The initial virial parameter is $\alpha_\mathrm{{vir}}=2E_\mathrm{{kin}}/E_\mathrm{{grav}}=0.5$, which is in the range of observed values \citep{1992A&A...257..715F,2013ApJ...779..185K,2015ApJ...809..154H}. A steady state of turbulence will be reached as the simulation reaches two turbulent crossing times \citep{2010A&A...512A..81F,PriceFederrath2010}, which is when self-gravity is turned on ($t=0$). The turbulence driving is maintained during the further evolution. The simulation is then allowed to evolve till a star formation efficiency (SFE) of 5\% is reached, i.e., when 5\% of the total cloud mass has formed stars. During this evolution phase, the high-density regions (analogous to dense cores) produced by the turbulent shocks become gravitationally unstable and form star clusters \citep[see][for further description of the numerical setup]{2021MNRAS.507.2448M,2023MNRAS.518.5190M}.

\subsection{Multiplicity identification algorithm}
\label{sec: multiplicity algorithm}
To identify the binary pairs in the simulations, we follow the iterative technique used in \citet{2009MNRAS.392..590B}. In each iteration, we find the closest bound pair in the list of sink particles. This pair is replaced in the list by a binary object with a mass that is equal to the sum of the respective pair. The position and velocity of the binary object are given by the centre-of-mass position and velocity of the pair, respectively. In the next iteration, using the new list, we again search for the closest bound pair and replace them with an object of higher order, i.e., if the pair consists of two single objects, they are replaced by a binary object or if the pair comprises of a single object and binary object, they are replaced by a triple. The process of finding and replacing the closest bound pair is iterated till none of the objects in the set are bound anymore or the only possible combination is a quintuple. Systems of order higher than quadruples are rejected since high-order systems are highly unstable and they can easily decay dynamically. The output of the above algorithm is a list of multiple systems, i.e., a list consisting of singles, binaries, triples, and quadruples. From this list, we can calculate different multiplicity properties of the systems such as the eccentricity and semi-major axis \citep{2023ASPC..534..275O}. This method is already used in \citet{2021MNRAS.507.2448M} and \citet{2023MNRAS.518.5190M} to determine multiplicity properties. The eccentricity $e$ of the binary orbit is given by,
\begin{equation} 
e = \sqrt{1 + \frac{2\varepsilon h^2}{G^2\left(M_1+M_2\right)^2}}.
\label{eq: e_eq}
\end{equation}
Here $\varepsilon, h$, and $G$ are the specific energy (kinetic+gravitational), specific angular momentum, and the gravitational constant, respectively. $M_1$ and $M_2$ correspond to the mass of the two binary components. Using Eq.~\ref{eq: e_eq}, we calculate the eccentricity associated with the binaries in our simulations and analyse the eccentricity distribution and its dependence on the turbulent driving mechanism, binary separation, mass and mass ratio, which is presented next.

\section{Results}
\label{sec: results}
To study the binary statistics, we utilise a total of 28~simulations of star cluster formation that form 1362~sink particles (stellar objects) overall. The sink particles themselves represent star+disc systems. However, to make discussions lucid, we will refer to sink particles as stars here although they might represent an under-resolved binary since disc fragmentation is not fully resolved (see discussion on limitations in \S\ref{sec: limitation}).

Using the method described in \S\ref{sec: multiplicity algorithm}, we categorise the stars formed in each of the simulations into multiple systems. The simulations have the same numerical setup and initial conditions as described in \S\ref{sec: initial conditions}, except that they are characterised by different modes of turbulence driving. In the suite of simulations used here, there are 7~simulations (forming a total of 468~stars) with a purely compressive mode of turbulence driving (curl-free driving), 11~simulations (forming 445~stars) with purely solenoidal turbulence driving mode (divergence-free), and 10~simulations (forming 449~stars) with a natural mixture (equal power in compressive and solenoidal modes). In each of the three sets, all simulations have the same setup but have different turbulence realisations (random seed to generate numerical turbulence). For example, a different turbulence realisation is used in each of the 7~simulations with purely compressive driving, providing a statistically conclusive sample of stars. The number of stars formed depends on the type of turbulence driving \citep{2023MNRAS.518.5190M}. Therefore, in order to obtain a similar number of stars (i.e., similar statistics) from each of the three simulation models, the number of simulations performed for each of the models is different. For example, the model with purely compressive turbulence driving produces the highest number of stars in a single simulation, and hence requires the least number of simulations to reach a total of $\sim450$ stars.

\begin{figure*}
    \centering
    \includegraphics[width=\textwidth]{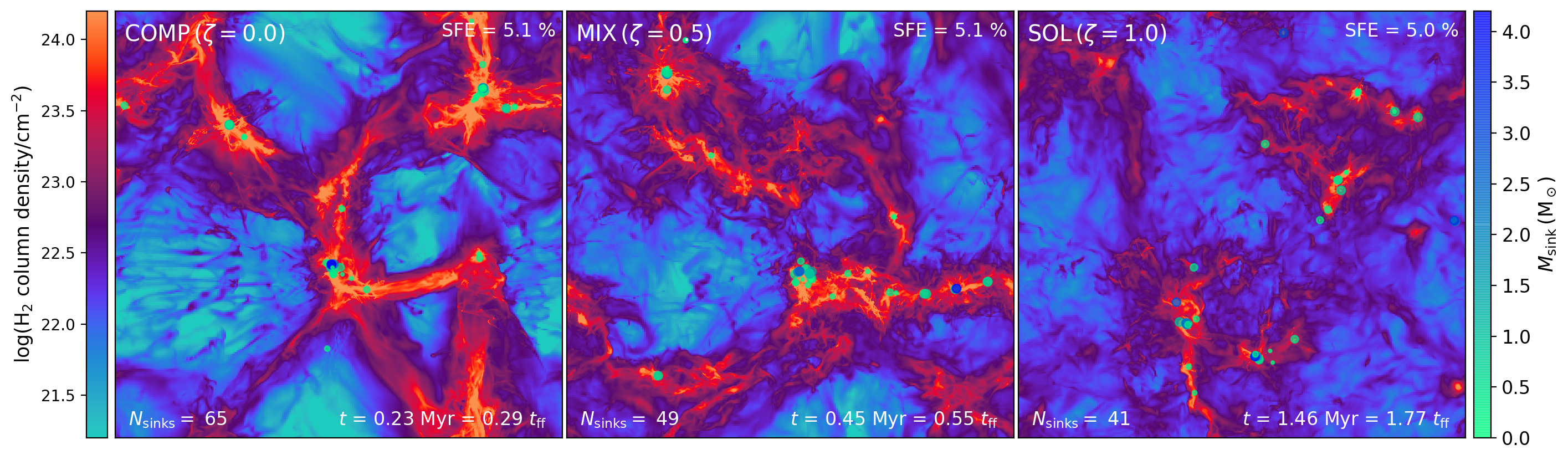}
    \caption{Column density maps (mass-weighted) of the purely compressive driving (COMP; left panel), mixed driving (MIX; middle panel), and solenoidal driving (SOL; right panel) simulations, at a star formation efficiency (SFE) of $5\%$. The circular markers correspond to the sink particle (star+disc system) positions, and the colour bar on the right represents the mass of the sink particles. The size of the markers is also scaled by the mass of the sink particles.}
    \label{fig:densmap}
\end{figure*}

\subsection{Effect of the mode of turbulence driving}
\begin{figure*}
    \centering
    \includegraphics[width=\textwidth]{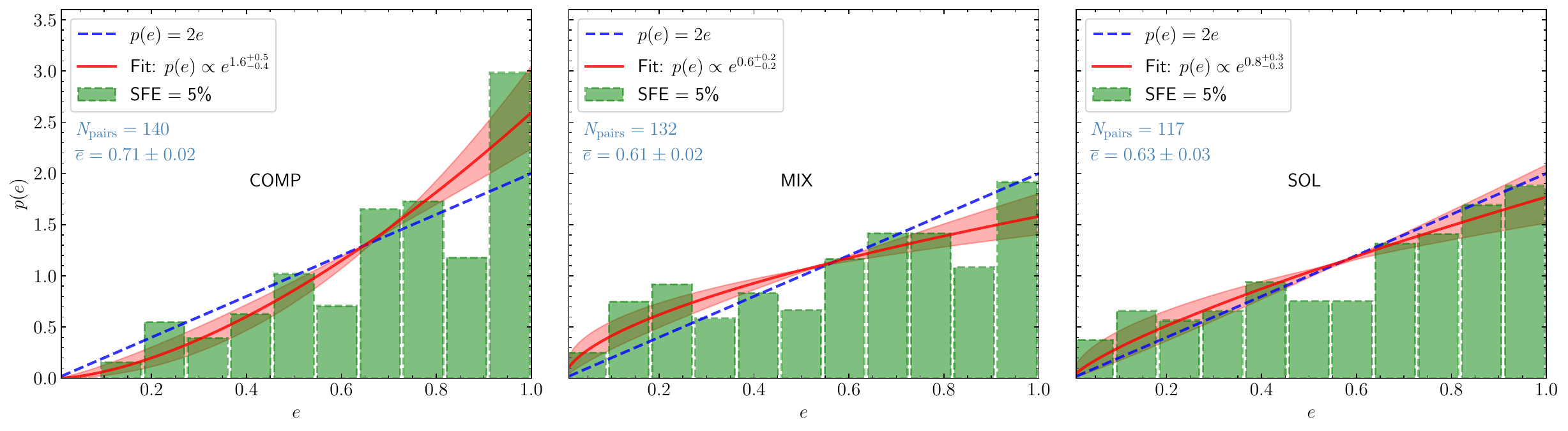}
    \caption{Eccentricity distribution of the bound pairs at SFE = 5\% for each of the three simulation models (from left to right: COMP, MIX, SOL). The solid curves represent power-law fits ($p(e) \propto e^{\alpha}$) to the distributions, and the dashed curves correspond to a thermal eccentricity distribution $p(e) \propto e^{1}$. The mean eccentricity is also indicated in the legends, which is higher for the COMP case than for MIX and SOL cases, which have similar values. The COMP case shows superthermal ($\alpha > 1$) eccentricity distribution whereas the MIX and SOL cases show subthermal to thermal ($\alpha \leq 1$) distributions. This demonstrates that the eccentricity distribution of binary stars depends on the turbulent driving mode of the parent cloud in which they are formed.}  
    \label{fig:e}
\end{figure*}

The mode of turbulence driving has a significant influence on the density distribution of a gas cloud and in turn on the stellar properties like the star formation rate \citep{2008ApJ...688L..79F,2010A&A...512A..81F,2011ApJ...730...40P,2011ApJ...743L..29H} and the initial mass function (IMF) \citep{2010A&A...516A..25S,2015MNRAS.449..662L,2023MNRAS.518.5190M}. To determine which mode of turbulence driving dominates the driving in a cloud, the turbulence driving parameter $b$ can be estimated. $b$ is the constant of proportionality in the relation between the standard deviation of gas density and Mach number, $\sigma_{\rho/\rho_0}=b\mathcal{M}$ \citep[see][for a detailed description of $b$]{2008ApJ...688L..79F,2010A&A...512A..81F}. Purely compressive driving has $b\sim1$ and purely solenoidal driving has $b\sim1/3$ \citep{2008ApJ...688L..79F}. The $b$ value of real clouds is between 1/3 and 1 \citep{FederrathEtAl2016,MenonEtAl2021,ShardaEtAl2022,DhawalikarEtAl2022,GerrardEtAl2023}, where a value of $b\sim0.4$ represents a natural mixture of the two driving modes \citep{2008ApJ...688L..79F,2010A&A...512A..81F}. Fig.~\ref{fig:densmap} presents the density structure in three simulations that have the same numerical setup and turbulence realisation but have different turbulence driving modes -- from left to right: purely compressive (COMP), natural mixture (MIX) and purely solenoidal (SOL). The simulations are compared at a star formation efficiency SFE = 5\% (i.e., the ratio of stellar mass to initial cloud mass). The morphology of the star-forming regions in the three simulation models are different and the number of stars formed also varies between them, as discussed in detail in \citet{2023MNRAS.518.5190M}. Here we primarily focus on the eccentricities of the binary stars.

Fig.~\ref{fig:e} compares the eccentricity distributions between the three simulation models. The histograms are normalised such that the total area under the histogram is unity. In each of the eccentricity distributions, only the binary pairs that contain the primary star (highest mass star) in the corresponding multiple system are included. For example, in a triple system where a binary subsystem (secondary+tertiary) orbits a primary, only the bound pair that involves the primary is included in the distributions (the secondary+tertiary subsystem is excluded). With such a selection criterion, the COMP model yields a total of 140~pairs, the SOL model gives 132~pairs, and the MIX model yields 117~pairs. The power-law fits to the eccentricity distributions ($p(e) \propto e^{\alpha}$) are obtained by Monte-Carlo sampling where the red solid line represents the $50^{\mathrm{th}}$ percentile and the shaded region is bracketed by the $16^{\mathrm{th}}$ and $84^{\mathrm{th}}$. The dashed line corresponds to a thermal distribution with $p(e)=2e$. The mean eccentricities for the COMP, MIX, and SOL models are $\overline{e}=0.71\pm0.02, 0.61\pm0.02$, and $0.63\pm0.03$, respectively (see Tab.~\ref{tab:e_cal}). The $e$ distribution of the COMP model is superthermal ($\alpha\sim1.6$) while the $e$ distributions for the MIX and SOL models are in the subthermal-thermal range ($\alpha\sim0.6$ and $\alpha\sim0.8$, respectively), and are similar to each other ($\alpha$ values are within the 1-sigma variations of each other). We performed Kolmogorov–Smirnov (K–S) tests to quantify the differences between the distributions. The p-values obtained when comparing the distributions for COMP and MIX is 0.034, and that for COMP and SOL is 0.11. These p-values suggest that $p(e)$ for COMP is likely different from that of MIX or SOL. The p-value of 0.59 obtained from comparing MIX and SOL implies that the $e$ distributions for MIX and SOL are likely similar. This is expected because the turbulence driving parameter $b$ for MIX and SOL is similar (0.4 and 0.3, respectively), i.e., MIX and SOL yield statistically very similar distributions. The variations caused by randomness are likely comparable to that caused by the small change in $b$ between MIX and SOL. Overall, it can be inferred that the $e$ distribution has a considerable dependence on the turbulence driving mode (SOL or MIX on one hand vs.~COMP on the other). When compressive turbulence driving dominates, the $e$ distribution tends to be superthermal, while for solenoidal or mixed driving, it corresponds to a subthermal/thermal distribution.

The effect of the turbulent properties of the cloud on the eccentricity statistics will be investigated further in a follow-up work. Here we will mainly focus on the effects of the binary parameters including the binary separation and mass. For the purpose of increasing our sample size, from here on, all analyses will be carried out by compiling together data from the three simulation models, i.e, using the binary pairs from the COMP, MIX, and SOL simulations, yielding a total of 389~binaries. We justify this choice of adding all the models together, in the next paragraph.

The data compilation we obtain will be analogous to binary data obtained from star-forming clouds in the Milky Way since the initial conditions employed in the simulations are typical of low-mass star-forming regions in the Milky Way. Similar to the different modes of turbulence driving obtained in the simulations ($b\sim1$, $b\sim0.4$, and $b\sim0.3$), the extent to which a particular driving mode dominates varies between the clouds in the Milky Way (i.e., a range of $b$ values can be associated with the clouds). The binaries in the solar neighbourhood collected by observational surveys possibly originate from clouds with different turbulence properties. Since these surveys generally do not distinguish the binary systems based on the initial turbulence environment, combining our simulation models is reasonable. Combining all the simulations also ensures that we have more statistics than the individual cases.

\begin{figure}
    \centering
    \includegraphics[width=\columnwidth]{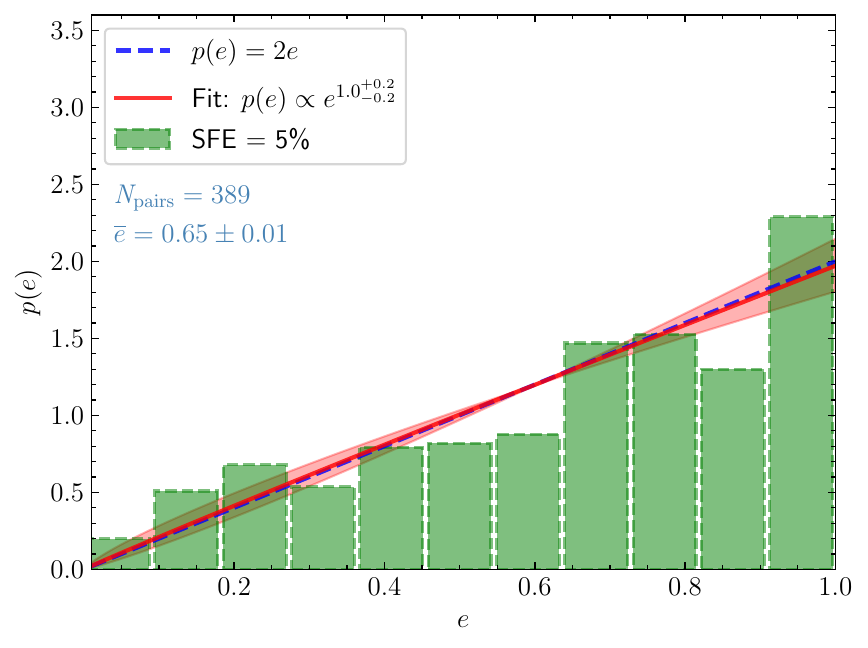}
    \caption{Eccentricity distribution at SFE = 5\% for the three simulation models compiled together. We see that the power-law fit (solid) agrees with a thermal distribution (dashed).}  
    \label{fig:e_all}
\end{figure}

Fig.~\ref{fig:e_all} presents the measured $p(e)$ when the data from the three simulation models are considered as a whole (i.e., plotting the data in the three panels of Fig.~\ref{fig:e} together). The resulting collection provides 389~binary pairs. The mean eccentricity is $\overline{e}=0.65\pm0.01$, denoting the presence of a relatively high fraction of highly eccentric orbits. It is evident from the figure that the fit to the distribution compares well with a thermal distribution of $p(e)=2e$. The thermal distribution here is simply a result of combining superthermal (COMP) and subthermal/thermal (MIX, SOL) distributions together. Fig.~\ref{fig:e} clearly shows that the physics of the turbulence gives rise to different distributions. This suggests that observations might be encouraged to look for non-thermal distributions, which might have been the results of different turbulence in different environments. The distribution corresponds to the simulation time at which SFE = 5\%. Since the star formation rate differs greatly between the three simulation models, the time at which the SFE = 5\% is reached varies between the simulations here. However, as compared to the field stars, the stars here are still very young, with even the oldest stars having a lifetime (time between star formation and end of simulation) of $< 1\, \mathrm{Myr}$. We examined the $p(e)$ at SFE = 2, 3, 5\%, combining the different simulation models together. The value of $\alpha$ measured for the three groups can be found in Tab.~\ref{tab:e_cal}. The p-value returned while comparing the $e$ distribution corresponding to SFE = 2\% and 3\% is 0.23. Comparisons between $p(e)$ at SFE = 2\% and SFE = 5\%, and $p(e)$ at SFE = 3\% and SFE = 5\% gives p-values of 0.69 and 0.93, respectively. The distributions are similar and approximately thermal for all SFE, i.e., at different points in time and star formation efficiency, implying that the overall form of the $e$ distribution is decided at the early stages of the cluster formation process.

Observational studies of the eccentricity in binaries generally find mean eccentricity values of < 0.5 \citep{1991A&A...248..485D,2006ApJ...651.1151A,2010ApJS..190....1R}, which is comparatively lower than the value calculated from our simulations ($\overline{e}\sim0.65$). However, the sample of binaries used in these studies have orbital periods of less than a few $10^{5}\, \mathrm{days}$, i.e., a semi-major axis of < $\sim100\, \mathrm{AU}$. Recently, it has been shown that the $e$ distribution is dependent on the separation between pairs or the semi-major axis \citep[e.g.,][]{2020MNRAS.496..987T,Hwang22}. Hence, the mean estimation will be sensitive to the selection scheme in the survey, i.e., whether the sample has a higher population of wide or close pairs. The binary pairs in our simulations represent a sample with a wide range of semi-major axes, varying from around 10~AU to $10^5\,\mathrm{AU}$, although we are underestimating the number of close binaries. \citet{2014MNRAS.442..285B} find a mean eccentricity of $0.33\pm0.02$ with a standard deviation of $0.28$ in their simulations, which have comparatively much higher maximum spatial resolution \citep[see also][]{2009MNRAS.392..590B,2012MNRAS.419.3115B,2019MNRAS.484.2341B,guszejnov23}. The mean eccentricity in \citet{2014MNRAS.442..285B} is almost a factor of $2$ less than our estimate, because a major fraction of their binary population has separations of around $10-20\, \mathrm{AU}$, while most of the binaries from this work have separations of $> 200\, \mathrm{AU}$. The binary statistics, including the eccentricity, are significantly different for different separation or semi-major axis ranges, which is discussed next.

\subsubsection{Semi-major axis distribution}

\begin{figure}
    \centering
    \includegraphics[width=\columnwidth]{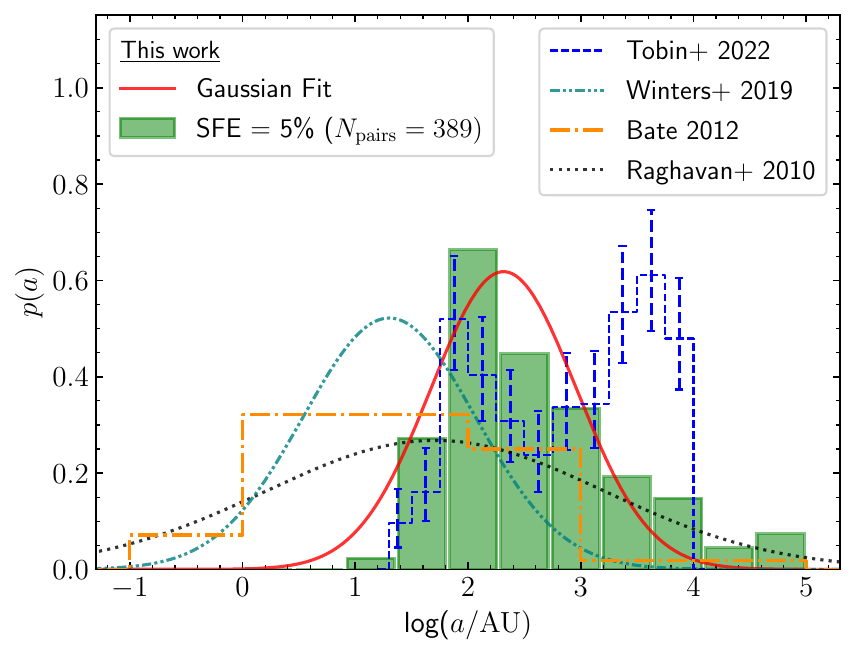}
    \caption{The distribution of the semi-major axis $a$ of the bound pairs from our simulations (histogram with solid edges). The solid line corresponds to a Gaussian fit with a peak at $a=206\,\mathrm{AU}$ and a standard deviation $\sigma_{\mathrm{log} (a/\mathrm{AU})}=0.98$. The dash-dotted histogram \citep[adapted from][]{2012MNRAS.419.3115B} represents the $a$ distribution (including the pairs in binary, triple, and quadruple systems) in the radiation hydrodynamical simulations of \citet{2012MNRAS.419.3115B}. The dotted curve represents $p(a)$ for solar-type binaries obtained by \citet{2010ApJS..190....1R}. The dash-double-dotted curve represents $p(a)$ for M-dwarf systems derived in the survey by \citet{2019AJ....157..216W}. The dashed histogram \citep[adapted from][]{2022ApJ...925...39T} corresponds to the bimodal $a$ distribution obtained for a binary sample in the Orion and Perseus molecular clouds combined together ($\sim400$ protostars), comprising mainly of Class~0 and Class~I objects.}  
    \label{fig:a_all}
\end{figure}

Fig.~\ref{fig:a_all} shows the binary semi-major axis formed in the three simulation models combined together, in comparison with the $a$ distributions collected from the radiation hydrodynamical simulations of \citet{2012MNRAS.419.3115B} (dash-dotted histogram in Fig.~\ref{fig:a_all}) and some observational surveys \citep{2010ApJS..190....1R,2019AJ....157..216W,2022ApJ...925...39T}. The peak of the distribution occurs at around 200~AU, although there exists a considerable number of pairs at separations greater than 1000~AU. The number of closer pairs (< 100~AU) is only a lower limit since fragmentation on small scales (at < $\sim200\, \mathrm{AU}$) is not resolved (see \S\ref{sec: limitation} for more details). The criterion for sink particle formation in our simulations does not allow new sink particles to form within the accretion radius of an existing sink particle (250~AU). Therefore, in our simulations, the pairs at separations less than 250~AU are probably the ones that formed at larger separations which later migrated to closer separations due to dynamical interactions \citep{2001AJ....122..432R,2002MNRAS.336..705B,2005A&A...439..565G,2010ApJ...725.1485O}. Although new sink particles cannot form within the accretion radius of an existing sink particle (in that case the gas is accreted onto the existing sink particle), there is no resolution-dependent restriction on how close two sink particles can come together as the trajectory of sink particles are decided by Lagrangian dynamics, i.e., the gravitational interaction between each other and with the gas. However, the gravitational potential is softened at distances of around 100~AU to prevent artificial collision between sink particles (see \S\ref{sec: limitation}).

\citet{2010ApJS..190....1R} find that, in the solar neighbourhood, the distribution of projected separation of binaries with solar-type primaries is Gaussian (on a logarithmic scale) with a peak at 50~AU (see dotted curve in Fig.~\ref{fig:a_all}) \citep[see also][]{1991A&A...248..485D, 1994ARA&A..32..465M}. For M-dwarf primaries in the solar neighbourhood, the distribution is Gaussian as well, but with a slightly lower peak of 20~AU \citep{2012ApJ...754...44J,2019AJ....157..216W}. The standard deviation of the distribution in \citet{2019AJ....157..216W} is $\sigma_{\mathrm{log}(a/\mathrm{AU}})\approx1.16$ (see dash-double-dotted curve in Fig.~\ref{fig:a_all}). The separation distribution of pre-main sequence (PMS) stars is found to be different from the distribution of the field stars with median separations estimated at a few hundred AU \citep{1994ARA&A..32..465M,2004A&A...427..651D,2008AJ....135.2526C,2013ApJ...768..110C,2016ApJ...818...73T}.

The VANDAM survey \citep{2016ApJ...818...73T} looked at a sample of 94~protostars comprising of Class~0 and Class~I objects within Perseus \citep{2009ApJ...692..973E} and obtained a bimodal separation distribution with a peak at $\sim75\, \mathrm{AU}$ and another peak at $\sim1000\, \mathrm{AU}$. \citet{2022ApJ...925...39T} find that the separation distribution is still bimodal with a peak at $\sim100\, \mathrm{AU}$ and a second peak at $\sim1000\, \mathrm{AU}$ after combining the samples from the Perseus \citep{2009ApJ...692..973E,2016ApJ...818...73T} and Orion molecular clouds \citep{2010A&A...518L.122F,2013ApJ...767...36S,2016ApJS..224....5F}, although the minimum projected separation that can be measured is $\sim20\, \mathrm{AU}$ (see dashed histogram in Fig.~\ref{fig:a_all}). \citet{2022ApJ...925...39T} suggest that the bimodal behaviour is mainly driven by the Class~0 protostars since the peak at $\sim1000\, \mathrm{AU}$ becomes insignificant for Class~I and more evolved protostars \citep[see also][]{kuruwita&haugboulle2023}. The age of the stars in our simulations lie in the range $10^3-10^5\, \mathrm{yr}$ which resembles the Class~0 and Class~I lifetimes of approximately $10^4$--$10^5\, \mathrm{yr}$ and a few $10^5 \mathrm{yr}$, respectively \citep{2006MNRAS.368..435F,2014prpl.conf..195D}. Therefore, the separation distribution in our simulations is more relatable to the distribution for young star-forming regions than that for field stars. However, the comparison of the position of the peak in the distribution with the observationally derived peaks should be done with caution because of the inherent limitations in the numerical resolution. This paper is focused on the relative differences in the binary statistics, particularly the variation in the eccentricity distribution for different separation and mass ranges, i.e., the relative change in the value of $\alpha$.

\begin{table*}
	\caption{Summary of eccentricity distribution results for different cases.}
	\label{tab:e_cal}
	\begin{tabular}{lcccccccc} 
	\hline
	\hline
    Driving & SFE & Range of $r_{\mathrm{pair}}$ & Range of $M_{\mathrm{sys}}$ & Range of $q$ & $N_{\mathrm{pairs}}$ & $\overline{e}$ & $\alpha$\\
    (1) & (2) & (3) & (4) & (5) & (6) & (7) & (8)\\
	\hline
    COMP & 5\% & FULL & FULL & FULL & 140 & $0.71\pm0.02$ & $1.6^{+0.5}_{-0.4}$ \\
    \hline
    MIX & 5\% & FULL & FULL & FULL & 132 & $0.61\pm0.02$ & $0.6^{+0.2}_{-0.2}$\\
    \hline
    SOL & 5\% & FULL & FULL & FULL & 117 & $0.63\pm0.03$ & $0.8^{+0.3}_{-0.3}$\\
    \hline
    ALL & 5\% & FULL & FULL & FULL & 389 & $0.65\pm0.01$ & $1.0^{+0.2}_{-0.2}$\\
    \hline
    ALL & 2\% & FULL & FULL & FULL & 249 & $0.63\pm0.02$ & $0.8^{+0.2}_{-0.2}$\\
    \hline
    ALL & 3\% & FULL & FULL & FULL & 316 & $0.66\pm0.01$ & $1.0^{+0.2}_{-0.2}$\\
    \hline
    ALL & 5\% & $\leq 1000\, \mathrm{AU}$  & FULL & FULL & 304 & $0.63\pm0.02$ & $0.8^{+0.2}_{-0.1}$\\
    \hline
    ALL & 5\% & $> 1000\, \mathrm{AU}$  & FULL & FULL & 85 & $0.73\pm0.02$ & $1.8^{+0.6}_{-0.4}$\\
    \hline
    ALL & 5\% & $1.5 \leq \mathrm{log}(r_{\mathrm{pair}}/\mathrm{AU}) < 2.5$  & FULL & FULL & 217 & $0.62\pm0.02$ & $0.7^{+0.2}_{-0.2}$\\
    \hline
    ALL & 5\% & $2.5 \leq \mathrm{log}(r_{\mathrm{pair}}/\mathrm{AU}) < 3.5$  & FULL & FULL & 133 & $0.67\pm0.02$ & $1.2^{+0.5}_{-0.3}$\\
    \hline
    ALL & 5\% & $3.5 \leq \mathrm{log}(r_{\mathrm{pair}}/\mathrm{AU}) < 4.5$  & FULL & FULL & 32 & $0.76\pm0.03$ & $2.5^{+1.5}_{-0.7}$\\
    \hline
    ALL & 5\% & FULL & $\leq 0.8\, \mathrm{M_\odot}$ & FULL & 77 & $0.70\pm0.03$ & $2.4^{+1.3}_{-0.9}$\\
    \hline
    ALL & 5\% & FULL & $> 0.8\, \mathrm{M_\odot}$ & FULL & 312 & $0.64\pm0.01$ & $0.8^{+0.1}_{-0.1}$\\
    \hline
    ALL & 5\% & FULL & FULL & $\leq 0.5$ & 203 & $0.65\pm0.02$ & $1.0^{+0.3}_{-0.2}$\\
    \hline
    ALL & 5\% & FULL & FULL & $> 0.5$ & 186 & $0.65\pm0.02$ & $1.0^{+0.3}_{-0.2}$\\
    \hline
    ALL & 5\% & $\leq 500\, \mathrm{AU}$ & FULL & $\leq 0.5$ & 117 & $0.64\pm0.02$ & $0.9^{+0.4}_{-0.3}$\\
    \hline
    ALL & 5\% & $\leq 500\, \mathrm{AU}$ & FULL & $> 0.5$ & 141 & $0.61\pm0.02$ & $0.6^{+0.2}_{-0.2}$\\
    \hline
    ALL & 5\% & $> 500\, \mathrm{AU}$ & FULL & $\leq 0.5$ & 86 & $0.67\pm0.03$ & $1.1^{+0.3}_{-0.3}$\\   \hline
    ALL & 5\% & $> 500\, \mathrm{AU}$ & FULL & $> 0.5$ & 45 & $0.75\pm0.03$ & $2.7^{+1.5}_{-1.0}$\\ 
    \hline
	\hline
	\end{tabular}
	\\
   \raggedright\textbf{Notes.} Column~1: MHD simulation model from which binaries are obtained ('ALL' means data from all simulation models are included, i.e., COMP, MIX, and SOL combined). Column~2: star formation efficiency (SFE) at which the calculations are made. Columns~3--5: selected range of separation ($r_{\mathrm{pair}}$), system mass ($M_{\mathrm{sys}}$), and mass ratio ($q$), with 'FULL' denoting the full range. Column~6: number of binaries that satisfy the constraints in Columns~1--5. Columns~7 and~8: mean eccentricity ($\overline{e}$) and power-law exponent ($\alpha$) in the eccentricity distribution, determined by fitting $p(e) \propto e^\alpha$.
\end{table*}

\subsection{Dependence on the binary separation}

\begin{figure*}
    \centering
    \includegraphics[width=0.7\textwidth]{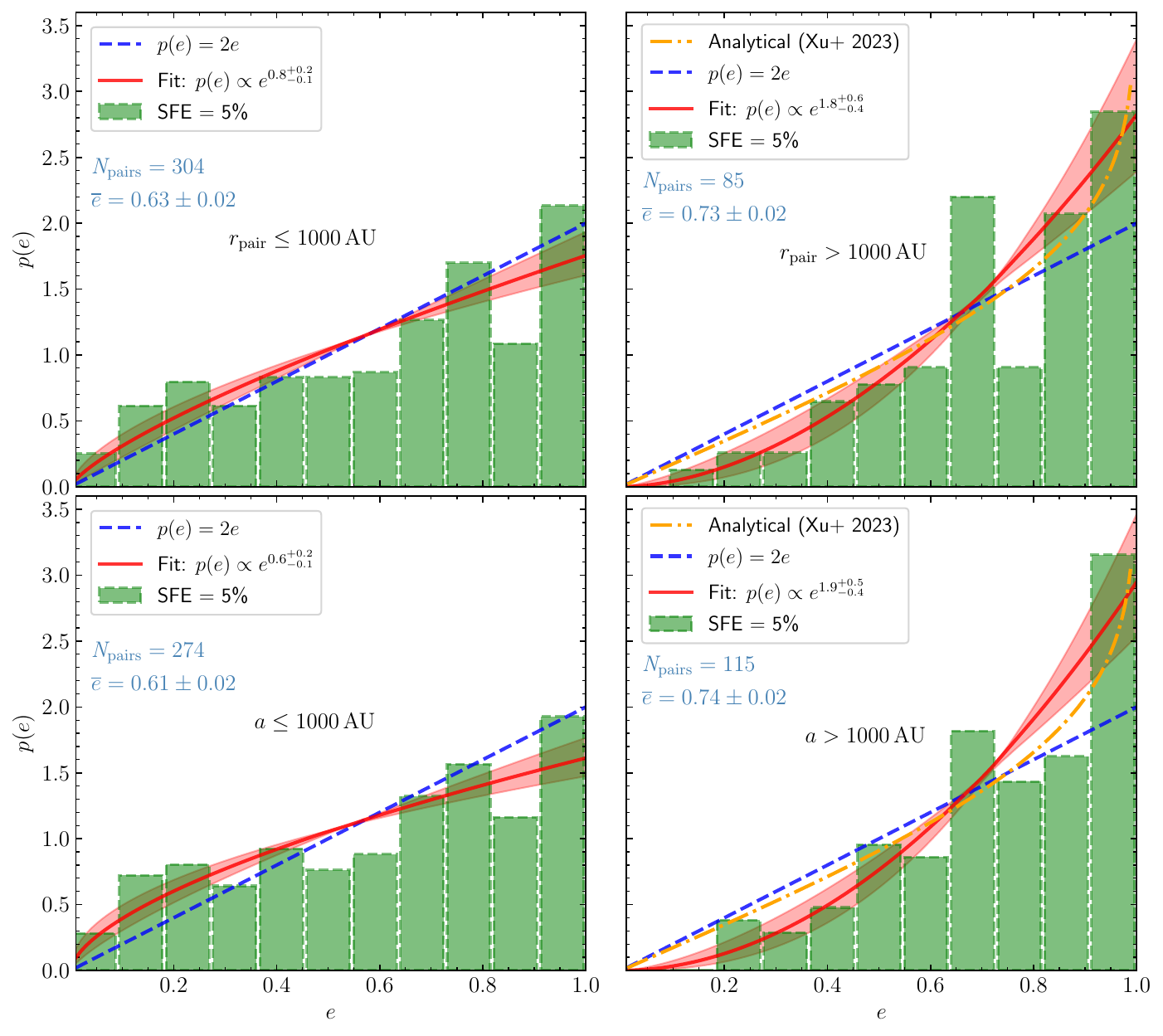}
    \caption{Top: The $e$ distribution at SFE = 5\% where the separation between the bound pairs is $r_{\mathrm{pair}} \leq 1000\, \mathrm{AU}$ (left) and $r_{\mathrm{pair}} > 1000\, \mathrm{AU}$ (right). Bottom: The $e$ distribution at SFE = 5\% where the semi-major axis of the bound pairs is $a \leq 1000\, \mathrm{AU}$ (left) and $a > 1000\, \mathrm{AU}$ (right). For the small separations, the distribution is sub-thermal, while the distribution is super-thermal for the large separations. This is consistent with observational findings by \citet{2020MNRAS.496..987T} and \citet{Hwang22}. The dash-dotted line represents the analytically derived $p(e)$ for wide binaries by \citet{Xu23}.}  
    \label{fig:e_abovebelowau}
\end{figure*}

The top row in Fig.~\ref{fig:e_abovebelowau} depicts the difference in $p(e)$ (at SFE = 5\%) between binary pairs with a separation of $r_{\mathrm{pair}}\leq1000\,\mathrm{AU}$ (close binaries) and those with $r_{\mathrm{pair}}>1000\,\mathrm{AU}$ (wide binaries). We see that wide binaries have a higher fraction of highly eccentric orbits. The mean eccentricities for the close and wide binaries are $\overline{e}=0.63\pm0.02$ and $\overline{e}=0.73\pm0.02$, respectively. It is clear from the fits that the distribution is in the subthermal to thermal range for close binaries ($\alpha\sim0.8$) and superthermal ($\alpha\sim1.8$) for wide binaries, which is consistent with the observational findings by \citet{2020MNRAS.496..987T} and \citet{Hwang22}. The p-value derived from the K-S test on the distributions is 0.013, which means that the distributions are most probably different. \citet{Xu23} propose that the superthermal nature of wide binaries is an outcome of star formation in the turbulent interstellar medium, where the scaling relations between the velocity differences and the initial separation of stars is regulated by turbulence. Their analytical derivation for $p(e)$ \citep[see Eq. 12 in][]{Xu23} agrees comparatively well with our $e$ distribution for wide binaries (see top-right and bottom-right panels in Fig.~\ref{fig:e_abovebelowau}). 

The bottom row presents the $e$ distribution when the distinction is based not on the separation but on the semi-major axis $a$. The p-value obtained from the K-S test is $10^{-3}$, which means that we can reject the hypothesis that the distributions are derived from the same underlying distribution. We see that the trend is similar to that in the top row with $\overline{e}=0.61\pm0.02$ and $\alpha\sim0.6$ for pairs with $a\leq1000\, \mathrm{AU}$ and $\overline{e}=0.74\pm0.02$ and $\alpha\sim1.9$ for binaries with $a>1000\, \mathrm{AU}$. It is possible that in a highly eccentric binary that has a large semi-major axis, the secondary (lower-mass star) may be close to the periastron (closest point) when the separation is measured and hence the separation between the components may be small at that instant of time, i.e., large $a$ but small $r_{\mathrm{pair}}$. However, since the velocity is low close to the apastron (furthest point) and high close to the periastron, at any time instance, it is more probable for the components to be at large separations in the case of wide pairs (large $a$). Hence the corresponding distributions in the top and bottom rows are statistically similar, i.e., the separation and semi-major axis are analogous when analysing the $e$ distribution \citep[see also][]{2011ApJ...733..122D,2022ApJ...933L..32H}. 

The mean eccentricity is found to be increasing on increasing the separation range by an order of magnitude with $\overline{e}=0.62\pm0.02$ in the range $1.5 \leq \mathrm{log}(r_{\mathrm{pair}}/\mathrm{AU}) < 2.5$, $\overline{e}=0.67\pm0.02$ in the range $2.5 \leq \mathrm{log}(r_{\mathrm{pair}}/\mathrm{AU}) < 3.5$, and $\overline{e}=0.76\pm0.03$ in the range $3.5 \leq \mathrm{log}(r_{\mathrm{pair}}/\mathrm{AU}) < 4.5$. The distributions also transition from subthermal ($\alpha\sim0.7$) in $1.5 \leq \mathrm{log}(r_{\mathrm{pair}}/\mathrm{AU}) < 2.5$ to a thermal distribution ($\alpha\sim1.2$) in $2.5 \leq \mathrm{log}(r_{\mathrm{pair}}/\mathrm{AU}) < 3.5$, and then to a superthermal distribution ($\alpha\sim2.5$) in $3.5 \leq \mathrm{log}(r_{\mathrm{pair}}/\mathrm{AU}) < 4.5$ (see Tab.~\ref{tab:e_cal}). This shows that the type of distribution can be different even when the separation ranges of interest differ by just an order of magnitude.

\citet{2020MNRAS.496..987T} finds that the $e$ distribution at $200-1000\, \mathrm{AU}$ separations is nearly thermal and at $< 200\, \mathrm{AU}$, the distribution has a clear shortage of highly eccentric pairs. They find that, at $> 10^3\, \mathrm{AU}$ separations, the distribution is slightly superthermal. \citet{Hwang22} find that the $p(e)$ is uniform at $\sim100\, \mathrm{AU}$, thermal at $\sim10^{2.5}-10^{3}\, \mathrm{AU}$ and superthermal at $> 10^3\, \mathrm{AU}$. Our findings are consistent with these observations.

We remind the reader that the actual lower error limits in the mean calculations and $\alpha$ estimates might be even lower. This is because we are probably underestimating the number of low eccentricities due to the inherent numerical limitations (see \S\ref{sec: limitation}). The underestimation is not taken into account while calculating error limits because an extensive resolution study is needed to quantify the extent of underestimation, which would require a separate study altogether. The underestimation becomes particularly significant for close binaries, e.g., in the above discussed binary range $1.5 < \mathrm{log}(r_{\mathrm{pair}}/\mathrm{AU}) < 2.5$.

\begin{figure}
    \centering
    \includegraphics[width=\columnwidth]{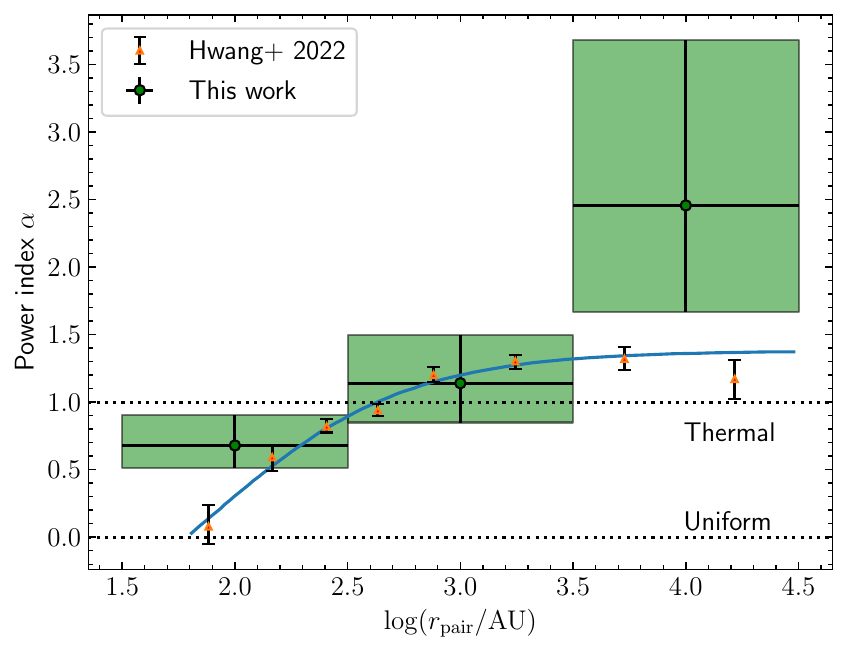}
    \caption{The power-law index $\alpha$ of the $e$ distribution at SFE = 5\% as a function of the separation. The circular markers within the rectangular boxes represent the $50^{\mathrm{th}}$ percentile estimate of $\alpha$ in our simulations in the separation range denoted by the width of the boxes. The height of the boxes represents the $16^{\mathrm{th}}$ to $84^{\mathrm{th}}$ percentile range of $\alpha$. The triangular markers correspond to the $\alpha$ values measured in \citet{Hwang22} at different binary separations, and the solid curve represents the fit to their data points.}  
    \label{fig:e_pow_vs_sep}
\end{figure}

Fig.~\ref{fig:e_pow_vs_sep} compares the value of $\alpha$ obtained in the three separation ranges ($1.5 \leq \mathrm{log}(r_{\mathrm{pair}}/\mathrm{AU}) < 2.5$, $2.5 \leq \mathrm{log}(r_{\mathrm{pair}}/\mathrm{AU}) < 3.5$, and $3.5 \leq \mathrm{log}(r_{\mathrm{pair}}/\mathrm{AU}) < 4.5$) with the measurements of $\alpha$ as a function of the binary separation by \citet{Hwang22} using the {\it Gaia} data. The $\alpha$ estimates in the ranges $1.5 \leq \mathrm{log}(r_{\mathrm{pair}}/\mathrm{AU}) < 2.5$ and $2.5 \leq \mathrm{log}(r_{\mathrm{pair}}/\mathrm{AU}) < 3.5$  agree well with the measurements of \citet{Hwang22}. However, our simulations produce a higher value of $\alpha$ in the range $3.5 \leq \mathrm{log}(r_{\mathrm{pair}}/\mathrm{AU}) < 4.5$ as compared to that for the wide binaries with separations of $>  10^{3.5}\, \mathrm{AU}$ in \citet{Hwang22}. The excess of highly eccentric orbits is most likely because the stars in our simulations represent a very young cluster as compared to the ones probed by the {\it Gaia} observations. The value of $\alpha$ for the wide binaries is likely to be reduced with further evolution in the Galactic field as a result of physical processes like dynamical scatterings between binaries and passing stars and molecular clouds \citep{Mod23,Hamil23}.

\subsection{Dependence on the binary mass and mass ratio}

\begin{figure*}
    \centering
    \includegraphics[width=0.8\textwidth]{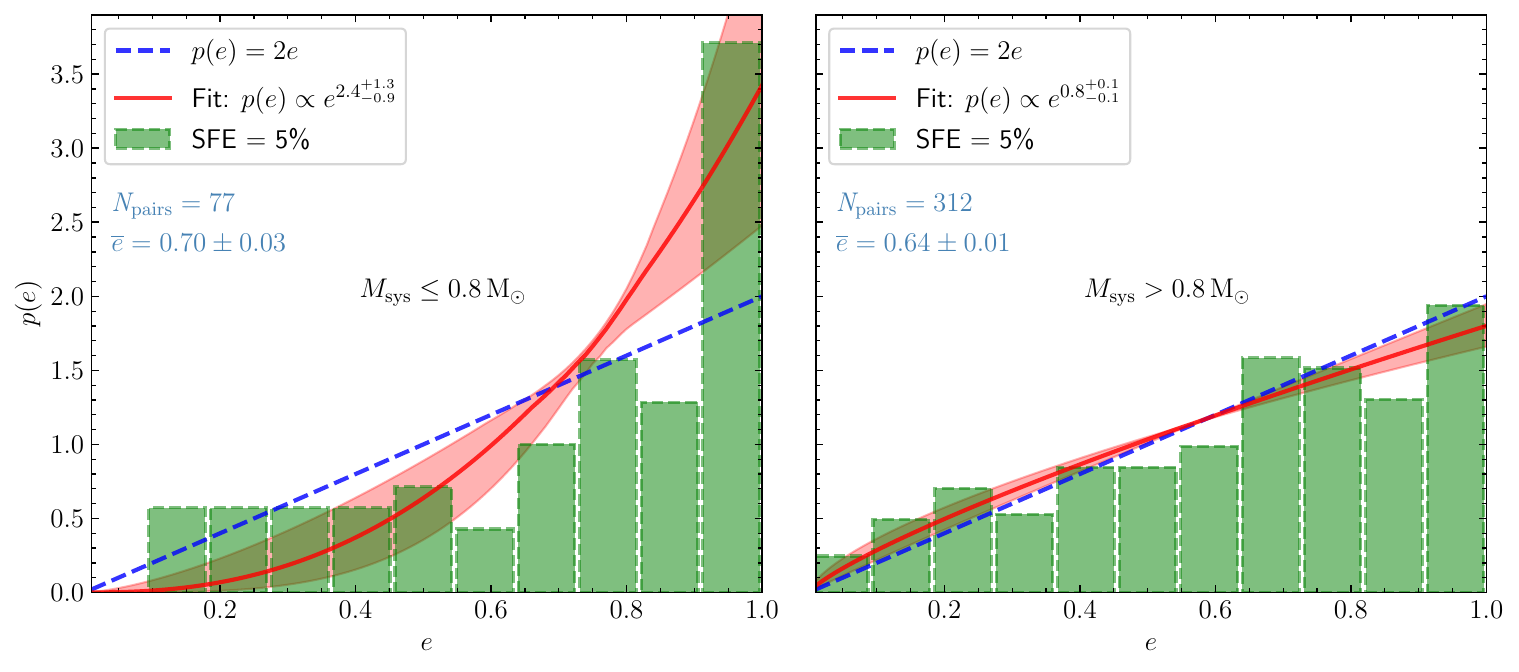}
    \caption{The $e$ distribution at SFE = 5\% for binaries with system masses $M_{\mathrm{sys}} \leq 0.8\, \mathrm{M_\odot}$, i.e., binaries with mainly M-dwarfs and later type stars  (left) and $M_{\mathrm{sys}} > 0.8\, \mathrm{M_\odot}$, i.e., binaries with mainly solar and earlier type stars (right). The low-mass systems ($M_{\mathrm{sys}} \leq 0.8\, \mathrm{M_\odot}$) have highly superthermal $e$ distribution while the relatively higher-mass systems have subthermal/thermal distribution.}  
    \label{fig:e_diff_sysmass_mdwarfs}
\end{figure*}

Fig.~\ref{fig:e_diff_sysmass_mdwarfs} shows $p(e)$ for low-mass binary systems ($M_{\mathrm{sys}} \leq 0.8\, \mathrm{M_\odot}$, left panel) and high-mass systems ($M_{\mathrm{sys}} > 0.8\, \mathrm{M_\odot}$, right panel), where $M_{\mathrm{sys}} = M_1 + M_2$. A large fraction of the low-mass binaries have highly eccentric orbits ($e>0.9$) with $\overline{e}=0.70\pm0.03$. The p-value obtained from the K-S test on the distributions is 0.022, and therefore we can say that the two distributions are different with 98\% confidence. Further, the low-mass systems have a highly superthermal distribution ($\alpha\sim2.4$) while the distribution for the high-mass systems is almost thermal ($\alpha\sim0.8$, $\overline{e}=0.64\pm0.01$), i.e., $p(e)$ is dependent on the total mass of the binary. This suggests that the form of the $e$ distribution obtained in observations would vary depending on the mass range of binaries selected in the survey \citep[see also][]{2011ApJ...733..122D}.

\begin{figure*}
    \centering
    \includegraphics[width=0.7\textwidth]{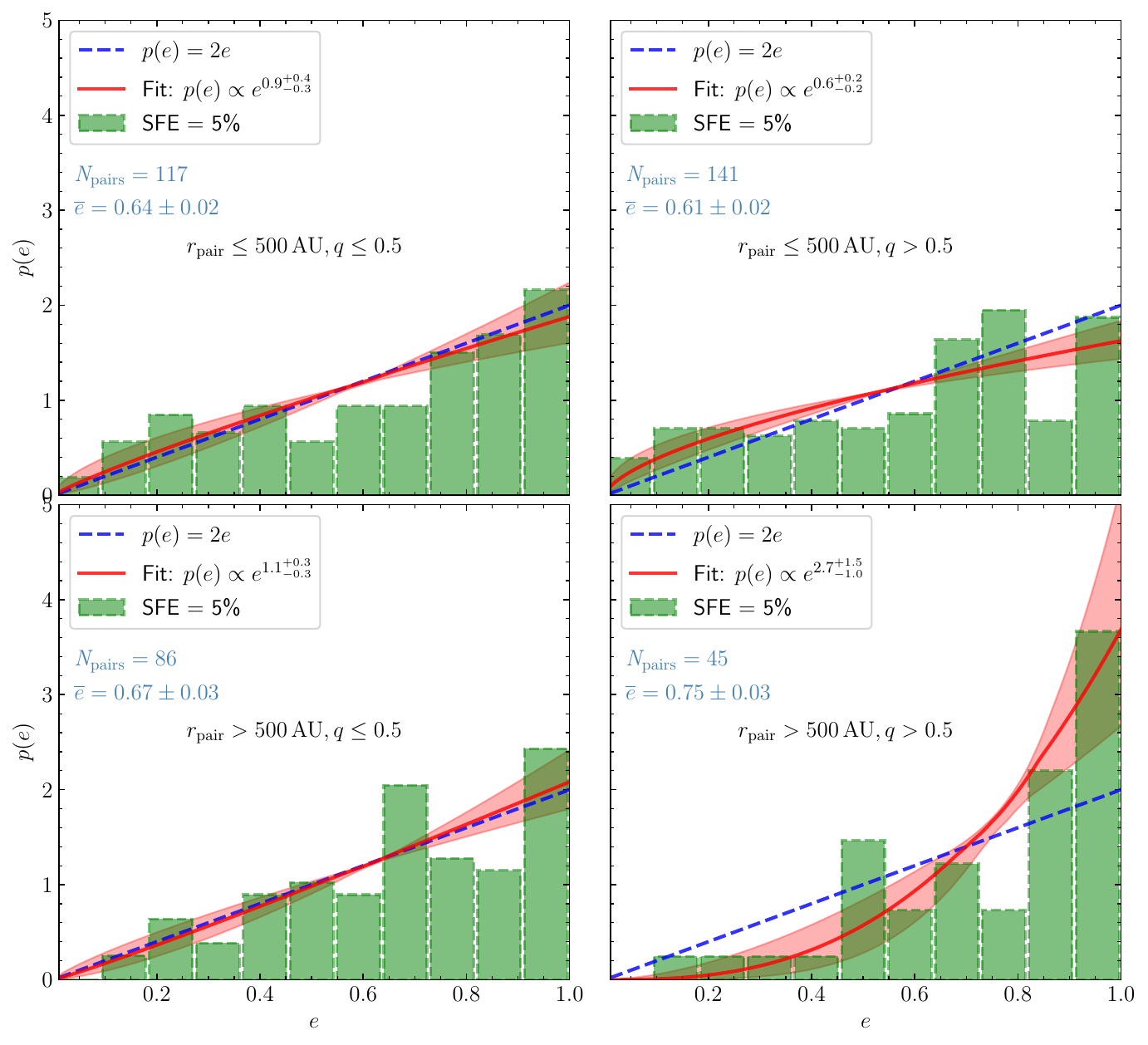}
    \caption{Top: The $e$ distribution at SFE = 5\% for binary systems with $r_{\mathrm{pair}} \leq 500\, \mathrm{AU}$ and $q$ value given by $ q \leq 0.5$ (left) and $q > 0.5$ (right). Bottom: The $e$ distribution at SFE = 5\% for binaries with $r_{\mathrm{pair}} > 500\, \mathrm{AU}$ and $q$ value given by $ q \leq 0.5$ (left) and $q > 0.5$ (right).}  
    \label{fig:e_qratio_wideandshort}
\end{figure*}

We compared $p(e)$ for binaries with a mass ratio ($q = M_2/M_1$, where $M_2 < M_1$) of $q \leq 0.5$ and those with $q > 0.5$ and find that the distributions for both groups are similar (p-value = 0.96) and represent thermal distributions, suggesting that the form of the $e$ distribution is generally independent of the $q$ value (see Tab.~\ref{tab:e_cal}). However, when looked at on different separation scales, the $q$ value does seem to influence $p(e)$. The top row in Fig.~\ref{fig:e_qratio_wideandshort} compares the $p(e)$ for binaries with $q \leq 0.5$ (left panel) and $q > 0.5$ (right panel) in the separation range $r_{\mathrm{pair}} \leq 500\, \mathrm{AU}$. The distribution is thermal ($\alpha\sim0.9$, $\overline{e}=0.64\pm0.02$) in the former and subthermal in the latter ($\alpha\sim0.6$, $\overline{e}=0.61\pm0.02$). However, a K-S test returns a p-value of 0.39, which means it is possible that the distributions are similar. The opposite trend can be seen in the bottom panel, where the same comparison is made, but for binaries with $r_{\mathrm{pair}} > 500\, \mathrm{AU}$. For $q \leq 0.5$ (left panel), we have a thermal distribution ($\alpha\sim1.1$, $\overline{e}=0.67\pm0.03$), while for $q > 0.5$ (right panel), the distribution is highly superthermal ($\alpha\sim2.7$, $\overline{e}=0.75\pm0.03$). The p-value obtained from the K-S test is 0.097, i.e., we can say with 90\% confidence that the distributions are different. The features seen in Fig.~\ref{fig:e_qratio_wideandshort} suggest again that the properties related to the eccentricity distribution are dependent on the separation range. The superthermal distribution obtained for the wide binaries with high mass ratios agrees with the observational finding in \citet{2022ApJ...933L..32H}.

\begin{figure}
    \centering
    \includegraphics[width=\columnwidth]{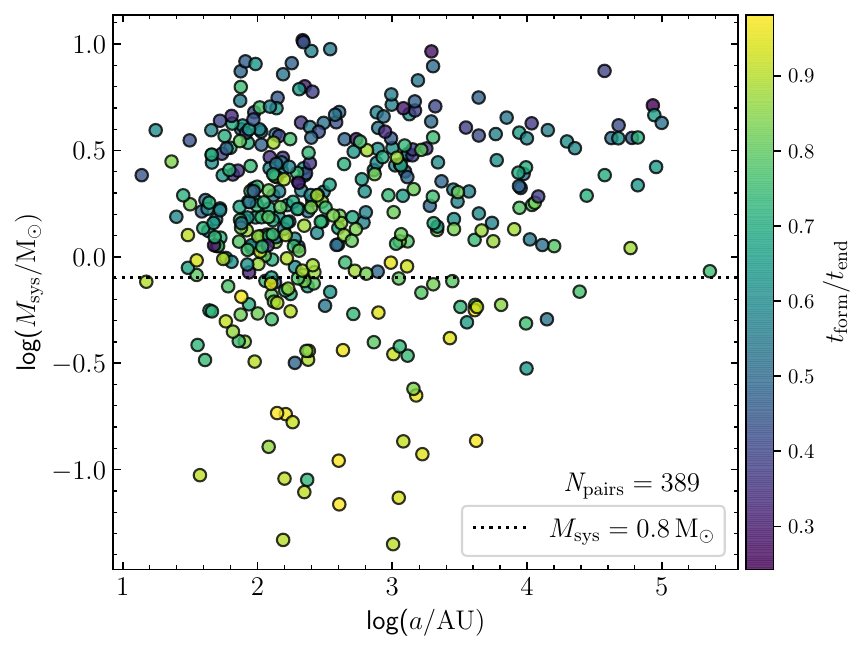}
    \caption{System mass ($M_{\mathrm{sys}}$) vs.~semi-major axis ($a$) at SFE = 5\%. The dashed line represents $M_{\mathrm{sys}}=0.8\, \mathrm{M_\odot}$. The marker colour is scaled to the time of formation of the binary ($t_{\mathrm{form}}$) in units of the simulation end time ($t_{\mathrm{end}}$).} 
    \label{fig:e_and_sys_mass_vs_a}
\end{figure}

Fig.~\ref{fig:e_and_sys_mass_vs_a} shows a scatter plot of the system mass $M_{\mathrm{sys}}$ as a function of the semi-major axis, $a$. The markers are colour-coded based on the time of formation (in units of the total simulation time) of the primary star in the pair. It is clear that the pairs with high system masses form in the early stages of cluster formation. These pairs also fall under a wide range of semi-major axis values ranging from around $10\, \mathrm{AU}$ to a few $10^5\, \mathrm{AU}$. The low-mass systems on the other hand form towards the last stages of the cluster formation and they lie mostly in the $100-1000\, \mathrm{AU}$ range. This, along with the finding that low-mass systems have an $e$ distribution that is different to that of high-mass systems (see Fig.~\ref{fig:e_diff_sysmass_mdwarfs}), suggests that different mechanisms may be responsible for binary formation on different mass scales, which is discussed next.

\section{Discussion}
\label{sec: discussion}
\subsection{Eccentricity distribution and binary formation mechanism}

In the following, we discuss several binary formation mechanisms, ranging from large-scale (cloud) to small-scale (disc) fragmentation.

\subsubsection{Fragmentation on cloud scales (fragmentation by turbulence-induced density fluctuations)}
Star formation is typically initiated by the turbulent shocks prevalent in molecular clouds. These shocks induce overdensities in the gas cloud, which, when they become gravitationally unstable, lead to the birth of stars \citep{2018ApJ...854...88R,2018MNRAS.480.3916M}. 
When a pair of stars (originating from two individually collapsing overdensities) are gravitationally bound at birth, they form as a binary. 
Due to the log-normal distribution of gas density in turbulent environments (e.g., \citealt{1994ApJ...423..681V,1997MNRAS.288..145P,2007ApJ...658..423K,2007ApJ...665..416K,2008ApJ...688L..79F,2013MNRAS.436.1245F,2013MNRAS.430.1880H,2015MNRAS.448.3297F,2019A&A...628A.112K,2022MNRAS.514..957S}), these overdensities vary significantly in size and mass. This variation gives rise to diverse stellar formations, including binary systems with different separations and system masses. Since the overdensities are produced by supersonic turbulent shocks, the binaries formed by pairing-up of individually collapsing overdensities will generally have separations comparable to the supersonic and transonic scales, i.e., $\gtrsim 0.1\, \mathrm{pc}$ \citep{2021NatAs...5..365F}.

Different mechanisms for the formation of wide binaries have been proposed in the literature 
\citep[e.g.,][]{2010MNRAS.404.1835K,2011MNRAS.415.1179M,2012Natur.492..221R,Pena16,2017MNRAS.468.3461T,2023MNRAS.521.4395L,2023ApJ...955..134R,2023MNRAS.521..866R}. \citet{Xu23} clearly addressed the role of turbulence 
and analytically derived that wide binaries ($\sim0.01 - 1\, \mathrm{pc}$) formed from turbulence-induced density fluctuations have superthermal $e$ distributions by considering that the initial velocity difference and separation of the binary components follow the turbulent gas velocity scaling. The superthermal $p(e)$ obtained for the high $r_{\mathrm{pair}}$ binaries in our simulations confirms their theoretical prediction (see Fig.~\ref{fig:e_abovebelowau}). We also find that the form of the $e$ distribution obtained for the simulations that employ purely compressive turbulence driving is different from that of the simulations with purely solenoidal driving or a natural mixture of turbulence driving modes. This indicates that the turbulent properties of the parent cloud play a critical role in the binary eccentricity statistics.  

\subsubsection{Fragmentation on core scales (fragmentation induced by turbulence, rotation, gravity, and outflows)}
As the formation of the star cluster proceeds, some of the overdensities fragment further (analogous to core fragmentation) due to the interplay between the inherent turbulence, rotation, and gravity \citep[e.g.,][]{1972MNRAS.156..437L,1992ApJ...388..392I,2002ApJ...576..870P,2008ApJ...684..395H,2009ApJ...702.1428H,2010ApJ...725.1485O,2012MNRAS.423.2037H,2012MNRAS.419.3115B,2015MNRAS.450.4137G,2023ASPC..534..275O}. In addition, protostellar outflow feedback promotes fragmentation by driving core-scale turbulence \citep{2014ApJ...790..128F,2021MNRAS.507.2448M,2021MNRAS.502.3646G,2022MNRAS.513.2100H}. The fragmentation on core scales would produce binaries with separations in the range of $\sim10^3\, \mathrm{AU}$ to $0.1\, \mathrm{pc}$ \citep{2023ASPC..534..275O}. On smaller scales, the turbulence transitions from transonic to subsonic \citep{2021NatAs...5..365F}, where gravity, thermal effects, magnetic fields, and rotation begin to dominate over turbulence. The formation of systems with very low masses and with close separations ($\lesssim500\, \mathrm{AU}$) at the time of their formation by turbulent fragmentation is relatively uncommon, as it is difficult to generate sufficiently strong overdensities through turbulent shocks on core scales, as those scales are intrinsically trans- to subsonic, at best very midly supersonic \citep{1978ApJ...220.1051E,1993MNRAS.263..701L,1999MNRAS.307..328C,2008MNRAS.389.1556B,2023ASPC..534..275O}.

The low-mass systems formed in the present simulations, which represent a relatively small population, have highly superthermal $e$ distributions (see Fig.~\ref{fig:e_diff_sysmass_mdwarfs}). These systems are found to have formed during the late stages of the cluster formation and have separations predominantly in the range $100-1000\, \mathrm{AU}$ (see Fig.~\ref{fig:e_and_sys_mass_vs_a}). These characteristics are unlike those of higher-mass systems, which form relatively early (likely via turbulence-induced density fluctuations in shocks; see above), and span wider separations of up to a few $10^5\, \mathrm{AU}$ (see Fig.~\ref{fig:e_and_sys_mass_vs_a}). The variation in the characteristics of $p(e)$ can be explained by the significant changes in the environment in which the binaries form, at early vs.~late stages or on large vs.~small scales. As stars continue to form in the shocked regions of gas, the local density increases and the local virial parameter decreases (more unstable gas) in cluster-forming regions due to the concentration of gas there. At late stages, the increase in the density and the increased influence of gravity allow the small collapsing overdensities to develop substructures with low Jeans length and low Jeans mass \citep{2008MNRAS.389.1556B}. The binaries that form by the gravitational fragmentation in these collapsing overdensities typically posses lower system masses and shorter separations compared to their counterparts that formed earlier through turbulent fragmentation. Numerical works \citep{2008MNRAS.389.1556B,2021MNRAS.507.2448M}, including the simulations presented here \citep[see Fig.~10 in][]{2021MNRAS.507.2448M}, have shown the existence of a negative correlation between the velocity at the time of formation and the final stellar mass, i.e., the low-mass (late-forming) stars have relatively high velocities as compared to the high-mass (early-forming) stars. This agrees with the argument that the low-mass stars that form in the late stages fragment out of the infalling gas \citep{2008MNRAS.389.1556B}. Such a negative correlation was also obtained in the recent observations of \citet{2023arXiv231204751W} \citep[see Fig.~10 \& 11 in][]{2023arXiv231204751W}. The superthermal nature of the binaries formed from gravitational fragmentation is possibly due to the preferential stretching of the binary orbit by the gravitational pull in the direction of the nearby sub-cluster. However, a detailed study would be required to test this scenario.

\subsubsection{Fragmentation on disc scales}
Fragmentation in extended discs \citep{2002MNRAS.332L..65B,2007A&A...466..943G,2007MNRAS.382L..30S,2011ApJ...730...32S,2012MNRAS.423.1896R,2015ApJ...800...72T} is likely to happen in our simulations. However, disc fragmentation on typical scales of $\lesssim 200\, \mathrm{AU}$ \citep[see reviews by][and the references therein]{2016ARA&A..54..271K,2020SSRv..216...70L} does not occur here, as these scales are not sufficiently resolved in our simulations (see \S\ref{sec: limitation}). We expect disc fragmentation \citep[e.g.,][]{1989ApJ...347..959A,1990ApJ...358..495S,1994MNRAS.269L..45B,2009MNRAS.392..413S,2011ApJ...730...32S,2016ARA&A..54..271K} to be an important binary formation mechanism \citep[see also][]{2019ApJ...883..140H} on these scales, favouring a uniform $e$ distribution \citep{2010ApJS..190....1R,2013ARA&A..51..269D,2017ApJS..230...15M,Hwang22,ceppi2024}, where very close binaries become circularised due to tidal effects. Thus, with higher numerical resolution, the $\alpha$ value in the short separation ranges is expected to be somewhat lower than suggested based on our current simulation results. However, since these simulations incorporate a large array of physical mechanisms and we carry out multiple such simulations to produce statistically significant data, achieving higher resolutions is not feasible at the moment, due to the associated computational cost, but remains a high priority for future work.    
    
\subsubsection{Dynamical interactions and gas friction/gas accretion}
Stars generally form in highly clustered environments \citep{2003ARA&A..41...57L}, and therefore dynamical interactions are a natural outcome of the star formation process. A considerable number of binaries undergo dynamical decay due to interactions with the gas or with other systems and migrate to closer separations in $\sim 100$~kyr \citep{2001AJ....122..432R,2002MNRAS.336..705B,2005A&A...439..565G,2016ApJ...827L..11O,2019ApJ...887..232L}. This explains why our simulations have a significant number of binaries with separations in the range $\sim10 - 200\, \mathrm{AU}$, even though fragmentation on these scales is not resolved in our simulations. A recent study by \citet{kuruwita&haugboulle2023} finds that core fragmentation and dynamical capture can produce a considerable number of low-mass close binaries (with disc-scale separations) via efficient in-spiral. During the migration, the eccentricity can change to a different value due to the dynamic nature of the orbital decay. 
For instance, the hydrodynamic drag forces cause eccentric orbits to become more circular \citep{2022MNRAS.513.5465S}. Similarly, initially close binaries can be widened due to interactions with other multiple systems and some of the initially wide binaries can even become unbound, since they are generally more weakly bound than close binaries.

\subsection{Future prospects}
An initial thermal eccentricity distribution \citep{Jean19} is frequently adopted in theoretical and numerical modelling of binary populations and star clusters \citep{kroupa1995,2021ascl.soft09029D}. However, adhering to a purely thermal eccentricity distribution can lead to inaccuracies in predicting the evolution of binary populations and their merger rates \citep{Gell19}. Our study provides physically justified and numerically tested initial eccentricity distributions of wide binaries for future studies. These distributions account for the complexities and variations introduced by different MHD turbulence properties in the surrounding medium. The superthermal eccentricity distribution of wide outer binaries in triple stellar systems also has important implications for the formation channel of black hole binary mergers \citep{2024arXiv240512270S}. \citet{2024arXiv240512270S} find that the outer eccentricity distribution can remain significantly superthermal for modestly hierarchical black hole triples.

The turbulence in our current simulations is weakly magnetized with $\mathcal{M_{\mathrm{A}}}\approx3$. 
Observations suggest variations of $\mathcal{M_{\mathrm{A}}}$ values in different molecular clouds \citep{Hu19}.
Earlier studies show that strong magnetic fields can play an important role in the formation of binaries, especially for the formation of massive close binaries \citep[e.g.,][]{2018MNRAS.479.2235L,2021MNRAS.508.3730H}. 
We will investigate the effect of magnetic fields on shaping the eccentricity distributions of binaries in our future work. 

\section{Limitations}
\label{sec: limitation}
\subsection{Numerical resolution}
Due to the limitations in the numerical grid resolution, the accretion discs are not fully resolved in our simulations. Further, to prevent artificial fragmentation due to resolution limitations, the sink particle formation criteria \citep{2010ApJ...713..269F} in our simulations does not allow the formation of a new sink particle within the accretion radius of an existing one (250~AU). Hence, the presence of close binaries ($\lesssim 200\, \mathrm{AU}$) in our simulations cannot be directly linked to fragmentation on those scales, e.g., disc fragmentation. These binaries are the ones that formed at relatively large separations initially and then decayed to small separations \citep{2001AJ....122..432R,2002MNRAS.336..705B,2005A&A...439..565G,2010ApJ...725.1485O}.

The resulting effect of the resolution limitation is that, although stars can form in extended discs in our simulations \citep{2002MNRAS.332L..65B,2007A&A...466..943G,2007MNRAS.382L..30S,2011ApJ...730...32S,2012MNRAS.423.1896R,2015ApJ...800...72T}, binary formation due to fragmentation on typical disc scales ($\sim100\,\mathrm{AU}$), particularly the formation of spectroscopic binaries (period of a few days), is underestimated. Spectroscopic binaries are found to have eccentricities close to zero \citep{2006ApJ...651.1151A} as a result of the tidal circularisation \citep{1977A&A....57..383Z,1992ApJ...395..259T}. Hence the fraction of pairs with low eccentricities is underestimated in our simulations.

\subsection{Gravitational softening}
When the separation between two star particles becomes very small, the gravitational acceleration becomes exceptionally high. As the distance approaches zero, the acceleration will go to infinity and the simulation timestep will tend to zero, which will stall the simulation. To overcome such a problem and also because the sink particle accretion radius is limited to the gas resolution, gravitational softening is introduced for distances shorter than a given softening radius (here equal to the sink particle accretion radius of $250\, \mathrm{AU}$). We utilise the version of spline softening \citep{2007MNRAS.374.1347P} used in \citet{2010ApJ...713..269F} to soften the gravitational interaction. The interaction is unaffected at distances greater than the softening radius, while at shorter distances, the acceleration smoothly approaches zero. Such a softening scheme may result in some of the orbits having artificial eccentricities if the binary spends time at separations shorter than the softening radius. The underestimation in the acceleration will be around a factor of $2$ at a distance of half of the softening radius \citep[see Fig.~1 in][]{2010ApJ...713..269F}.

The limitations in numerical resolution and the application of gravitational softening imply that the actual values of $\alpha$ are affected. However, the main objective of this study is to understand the relative change in $\alpha$, for example, when considering binaries in different separation and mass ranges. Therefore, at least the relative inferences from this paper still hold.

\section{Conclusions}
\label{sec: conclusions}
We use the star cluster formation simulations of \citet{2021MNRAS.507.2448M} and \citet{2023MNRAS.518.5190M} to investigate binary statistics, particularly the eccentricity distribution and its dependence on the binary separation, mass, and mass ratio. The simulations employ gravity, turbulence, magnetic fields, and protostellar feedback in the form of radiative heating and jets/outflows, and therefore include most of the main relevant physical ingredients for star and binary formation. The dataset comprises a total of 28~simulations, which together produce 1362~stars, of which 389 are binaries, allowing us to make statistically conclusive remarks. We find that

(1) the eccentricity distribution $p(e)$ is dependent on the mode of turbulence driving (see Fig.~\ref{fig:e}). The simulations with purely compressive driving produce $e$ distributions with the form $p(e) \propto e^\alpha$ with $\alpha>1$, while simulations with purely solenoidal and a natural mixture of driving modes have $\alpha<1$. Kolmogorov–Smirnov tests performed on the three simulation sets suggest that (with $\gtrsim 90\%$ confidence), the $e$ distribution obtained from the purely compressive driving simulations is different from that of simulations with purely solenoidal or a natural mixture of driving modes. This suggests that the turbulent properties of the cloud could play a significant role in shaping the $e$ distribution. We conclude that this is because the turbulence regulates the cloud dynamics including the density and velocity statistics of the gas from which the binaries emerge, and therefore influences their orbital parameters as well. 

(2) $p(e)$ is thermal ($\alpha\sim1$) when data from all the simulations are compiled together (see Fig.~\ref{fig:e_all}). Recent observational surveys also find that the $e$ distribution is thermal for broad separation ranges comparable to that in our simulations ($10-10^5\, \mathrm{AU}$). It is likely that these surveys include binary stars that formed in different turbulent conditions and environments, including different turbulence driving modes. It is also worth noting that the binary systems included in the surveys generally have undergone significant dynamical evolution in comparison to the young systems in our simulations, 
which tends to 
drive the eccentricity distribution towards thermal \citep[e.g.,][]{Hamil23}.

(3) $p(e)$ is similar with $\alpha\sim1$ at different points in time in our simulations where the cluster formation process is followed up to around $1\, \mathrm{Myr}$ (see Tab.~\ref{tab:e_cal}). This shows that the overall form of the $e$ distribution is imprinted in the early stages of the formation process, and small-scale dynamical interactions (within a cluster) do not significantly influence the $\alpha$ value.

(4) $p(e)$ is dependent on the binary separation or the semi-major axis, where wide binaries ($r_{\mathrm{pair}} > 1000\, \mathrm{AU}$ or $a > 1000\, \mathrm{AU}$) have superthermal distributions with $\alpha\sim2$ as compared to binaries with $r_{\mathrm{pair}} \leq 1000\, \mathrm{AU}$ or $a \leq 1000\, \mathrm{AU}$, which have subthermal distributions with $\alpha<1$ (see Fig.~\ref{fig:e_abovebelowau}). K-S tests suggest that the hypothesis that the two distributions are the same can be ruled out. The value of $\alpha$ transitions from subthermal to thermal, to superthermal, with increasing binary separation (see Tab.~\ref{tab:e_cal}), which concurs with the recent observational findings of e.g., \citet{2020MNRAS.496..987T}, \citet{Hwang22} from {\it Gaia} (see Fig.~\ref{fig:e_pow_vs_sep}).
We also find that the birth eccentricity distribution of wide binaries is more superthermal than the observed one in the Galactic field.

(5) $p(e)$ also depends on the system mass of the binary with low-mass systems ($M_{\mathrm{sys}} \leq 0.8\, M_\odot$) producing highly superthermal distributions  ($\alpha\sim2.4$), and higher-mass systems ($M_{\mathrm{sys}} > 0.8\, M_\odot$) producing subthermal to thermal $e$ distributions ($\alpha\sim0.8$) (see Fig.~\ref{fig:e_diff_sysmass_mdwarfs}). Based on a K-S test, we can say with 98\% confidence that the two distributions are different.

(6) $p(e)$ is independent of the binary mass ratio ($q$) when looked at in the whole separation range (see Tab.~\ref{tab:e_cal}). However, for relatively close binaries (here $r_{\mathrm{pair}} \leq 500\, \mathrm{AU}$), the distribution is thermal ($\alpha\sim0.9$) when $q\leq0.5$ and subthermal ($\alpha\sim0.6$) when $q>0.5$ (see Fig.~\ref{fig:e_qratio_wideandshort}). However, a K-S test suggests that there is a 40\% chance that the two distributions are similar. In the separation range $r_{\mathrm{pair}} > 500\, \mathrm{AU}$, the distribution is thermal ($\alpha\sim1.1$) when $q\leq0.5$ and highly superthermal ($\alpha\sim2.7$) when $q>0.5$, i.e., an opposite trend in the change of $\alpha$ as compared to that for close binaries ($r_{\mathrm{pair}} \leq 500\, \mathrm{AU}$). We find that the above distributions are different with 90\% confidence based on a K-S test.
Our finding is consistent with the observational finding on the highly superthermal eccentricity distribution of wide twin binaries 
\citep{2022ApJ...933L..32H}.

Our study suggests that the often adopted thermal eccentricity distribution \citep{Jean19} may not always be valid or appropriate, and properties of the cloud in which stars are formed have a direct influence on the form of the eccentricity distribution (see Tab.~\ref{tab:e_cal}). The dependence of the eccentricity distribution on the binary separation and mass indicates that different mechanisms are responsible for binary formation on different scales \citep[see also][]{2009MNRAS.392..590B,2012MNRAS.419.3115B,2014MNRAS.442..285B,2019MNRAS.484.2341B,guszejnov23}. Our results suggest that wide binaries ($r_{\mathrm{pair}} \gtrsim 1000\, \mathrm{AU}$) predominantly result from turbulence-induced density fluctuations \citep{Xu23} and turbulent fragmentation in cores \citep{2002ApJ...576..870P,2005ApJ...630..250K,2008ApJ...684..395H,2009ApJ...702.1428H,2010ApJ...725.1485O,2012MNRAS.423.2037H,2012MNRAS.419.3115B,2015MNRAS.450.4137G,2019A&A...628A.112K,2023ASPC..534..275O}. Our findings are consistent with the theoretical predictions of \citet{Xu23} that the superthermal nature of the eccentricity distribution of wide binaries is an outcome of the turbulent characteristics of the cloud in which they are born. In the intermediate separation range ($100 < r_{\mathrm{pair}}\, (\mathrm{AU}) < 1000$), turbulent fragmentation, gravitational fragmentation (\citet{2008MNRAS.389.1556B}; see also \citet{2018A&A...611A..88L}) and fragmentation in extended discs \citep{2002MNRAS.332L..65B,2007A&A...466..943G,2007MNRAS.382L..30S,2011ApJ...730...32S,2012MNRAS.423.1896R,2015ApJ...800...72T} may have a greater influence on the eccentricity distribution.

It is to be noted that, due to the highly dynamic nature of the star formation process, close binaries can form from initially wider binaries. The decrease of their separations can be caused by e.g., gas dynamical friction \citep{1999ApJ...513..252O,2019ApJ...887..232L,2023MNRAS.521..866R,rozner2024}, gas accretion \citep{2000MNRAS.314...33B}, and dynamical interactions in unstable multiple systems \citep{2001AJ....122..432R,2002MNRAS.336..705B,2005A&A...439..565G,2010ApJ...725.1485O}. Similarly, dynamical interactions can also widen the binary orbit \citep{2022ApJ...933L..32H}. On scales less than a few hundred AU, disc fragmentation \citep[e.g.,][]{1989ApJ...347..959A,1990ApJ...358..495S,1994MNRAS.269L..45B,2009MNRAS.392..413S,2011ApJ...730...32S,2016ARA&A..54..271K} may be a dominant mechanism leading to a uniform eccentricity distribution \citep{2010ApJS..190....1R,2013ARA&A..51..269D,2017ApJS..230...15M,Hwang22,ceppi2024}. However, disc fragmentation is not resolved in our simulations and hence on those scales, only binaries that formed from turbulent fragmentation, which then dynamically decay to short separations exist. 

We conclude that the properties of binaries, in particular their eccentricity, separation, total mass, and mass ratio, depend on the turbulence properties of the parent cloud from which they form.  
The statistical studies of binaries in young clusters provide valuable constraints on binary formation mechanisms and physically justified initial conditions for binary population synthesis.

\section*{Acknowledgements}
We thank the anonymous referees for their valuable remarks which improved the quality of the paper. C.F.~acknowledges funding provided by the Australian Research Council (Future Fellowship FT180100495), and the Australia-Germany Joint Research Cooperation Scheme (UA-DAAD). We further acknowledge high-performance computing resources provided by the Leibniz Rechenzentrum and the Gauss Centre for Supercomputing (grants~pr32lo, pr48pi and GCS Large-scale project~10391), the Australian National Computational Infrastructure (grant~ek9) and the Pawsey Supercomputing Centre (project~pawsey0810) in the framework of the National Computational Merit Allocation Scheme and the ANU Merit Allocation Scheme. The simulation software, \texttt{FLASH}, was in part developed by the Flash Centre for Computational Science at the University of Chicago and the Department of Physics and Astronomy at the University of Rochester. S.X. acknowledges support from grant NSF PHY-2309135 to the Kavli Institute for Theoretical Physics (KITP), the Aspen Center for Physics, which is supported by National Science Foundation grant PHY-2210452, Durand fund, a grant from the Simons Foundation (1161654, Troyer), and NASA ATP award 80NSSC24K0896. S.X. also acknowledges the inspiring discussion with the participants of the program on “Turbulence in Astrophysical Environments” at KITP. This work makes use of the yt-project \citep{yt} and colormaps in the CMasher package \citep{cmasher}.

\section*{Data Availability}
The data used in this article is available upon reasonable request to the authors.


\bibliographystyle{mnras}
\bibliography{Bibliography}






\bsp	
\label{lastpage}
\end{document}